\documentclass[12pt]{article}
\usepackage[dvips]{epsfig}
\usepackage{latexsym}
\usepackage[bf,footnotesize]{caption}	% Customize captions
\begin{document}
\setlength{\unitlength}{1mm}

\newcommand{\tBox} {\stackrel{\sim}{\stackrel{}{\Box}}}
\newcommand{\ba} {\begin{eqnarray}}
\newcommand{\ea} {\end{eqnarray}}
\newcommand{\be}{\begin{equation}}
\newcommand{\ee}{\end{equation}}
\newcommand{\n}[1]{\label{#1}}
\newcommand{\eq}[1]{Eq.(\ref{#1})}
\newcommand{\ind}[1]{\mbox{\tiny{#1}}}
\renewcommand\theequation{\thesection.\arabic{equation}}

\newcommand{\la}{\langle}
\newcommand{\ra}{\rangle}

\newcommand{\nn}{\nonumber \\ \nonumber \\}
\newcommand{\nl}{\\  \nonumber \\}
\newcommand{\pr}{\partial}
\renewcommand{\vec}[1]{\mbox{\boldmath$#1$}}

\title{
Quantum Radiation of Uniformly Accelerated Spherical Mirrors}
\author{\\
V. Frolov\thanks{e-mail: frolov@phys.ualberta.ca}
\  and 
D. Singh\thanks{e-mail: 
singh@phys.ualberta.ca}
\date{\today}}
\maketitle
\noindent 
{
\centerline{ \em
Theoretical Physics Institute, Department of Physics,} \\ 
\centerline{ \em University of Alberta, Edmonton, Canada T6G 2J1}
}
\bigskip

\begin{abstract} 
We study quantum radiation generated by a uniformly accelerated motion
of small spherical mirrors. To obtain Green's function for a scalar
massless field we use Wick's rotation. In the Euclidean domain the
problem is reduced to finding an electric potential in 4D flat space in
the presence of a metallic toroidal boundary. The latter problem is
solved by a separation of variables. After performing an inverse Wick's
rotation we obtain the Hadamard function in the wave-zone regime and
use it to calculate  the vacuum fluctuations and the vacuum expectation
for the energy density flux in the wave zone. 
\end{abstract}

\bigskip

\centerline{\it PACS number(s): 03.70.+k, 11.10.-z, 42.50.Lc}

\baselineskip=.6cm
 
\newpage

\section{Introduction}

In this paper we study quantum radiation generated by a uniformly
accelerated motion of a small spherical mirror. This is a special case
of the dynamical Casimir effect  \cite{DeWi:75}--\cite{BiDa:82}.   

There are several reasons that make studies of the dynamical Casimir
effect, that is the effect of vacuum polarization and particle creation
in the presence of moving boundaries, important.  For a special case
when boundaries are moving with constant acceleration, there is some
similarity of this problem with problems in a constant gravitational
field and Hawking effect. Schwinger \cite{Schw:93,Schw:94} suggestion
that the photon production associated with changes in the quantum
vacuum state in a system with collapsing dielectric bubble could be
responsible for sonoluminescence has generated a lot of publications
(see e.g. a review by Milton \cite{Milt:99}). System including mirrors
moving in an external gravitational fields were often used in different
gedanken experiments in studies of black holes. In particular,  Unruh
and Wald in their work devoted to the generalized second law of black
hole physics \cite{UnWa:82}, considered  a gedanken experiment in which
a box with mirrored walls filled with radiation is lowered
adiabatically toward a black hole. At any particular moment of time a
box is practically at rest in a static gravitational fields and hence
its boundaries have a non-vanishing constant acceleration. Unruh and
Wald demonstrated that the quantum radiation emitted by the mirror
boundaries during the adiabatic transition from initial to final
position of the box plays a key role in the understanding of the
mechanism providing the fulfillment of the generalized second law for
such processes. Moving mirrors  may also have an application in studies
of quantum radiation created by bubble formation during first order
phase transitions in the Early Universe.

In a 2 dimensional spacetime the dynamical Casimir effect is studied 
quite well \cite{Moor:70}--\cite{EORBZ:94}.  Much less results have
been obtained for the relativistic motion of mirrors in physical 4
dimensional spacetime.   Quantum radiation from a uniformly accelerated
plane mirror was considered in \cite{CaDe:77}. Quantum effects in the
presence of relativistic spherical mirrors which are expanding with
uniform acceleration  were studied in \cite{FrSe:79}--\cite{FrSe:80}. 
Ford and Vilenkin \cite{FoVi:82} have extended the plane mirror result
to include non-constant acceleration for the case when the acceleration
and its derivatives are small.

Recently the interest to the ``old'' problem of quantum effects in the
presence of accelerated mirrors increased. Partially this was
stimulated by attempts  to obtain a more detailed quantitative
description of  gedanken experiments with black holes for a more
realistic situation, when a 4D mirror has a finite size, is
semi-transparent and moves with an acceleration which not constant in
time. Since such a problem, especially in an arbitrary external
gravitational field is technically very complicated several
simplified models were considered. Anderson and Israel \cite{AnIs:99} calculated quantum fluxes from a spherical 4D mirror which is expanding or contracting with nearly uniform acceleration. Particle creation for adiabatic expansion or contraction of a spherical mirror and for an oscillating spherical cavity were discussed by Setare and Saharian \cite{SeSa:01a, SeSa:01b}. Quantum effects in a presence of an expanding or contracting (with constant acceleration) spherical semitransparent 4D mirrors were studied in \cite{FrSi:99}. Two-dimensional analogue of this problem for an arbitrary acceleration was discussed by Nicolaevici \cite{Nico:01}. Similar problem in a more general set-up was considered by Obadia and Parentani \cite{OdPa:01}. Energy flux and vacuum polarization created by a polarizable body of small size and moving with constant acceleration was calculated in  \cite{FrSi:00} 

Very recently systems with mirrors were used for study the problem of
the black hole entropy. Using a 2D moving mirror model Mukohyama and
Israel \cite{MuIs:00} introduced a notion of moving-mirror entropy
associated with temporarily inaccessible information about the future.
Page \cite{Page:01} argued that by surrounding a black hole by a
spherical reflecting shell (mirror) one can make the black hole entropy
less that its Bekenstein-Hawking value.

In this paper we continue studying the quantum effects in the presence
of a 4D moving mirror. We assume that a mirror has a spherical form and
it is uniformly accelerated.  As a model of such a system  we consider
a scalar massless field obeying the Dirichlet boundary conditions on
the surface of a tube generated by a uniformly accelerated motion (with
the acceleration $w$) of a sphere of radius $b$. The complete
information about quantum properties of the system is contained in the
Feynman Green's function. To obtain the latter we use Wick's rotation.
In the Euclidean domain the problem is reduced to finding an electric
potential in 4D flat space in the presence of a metallic toroidal
boundary. The problem we study  is greatly simplified by the presence
of high symmetry. As a result of this symmetry we can solve the problem
by a separation of variables by using the toroidal coordinates and
obtain a series representation for the Euclidean Green's function.
After performing an inverse Wick's rotation we obtain the Hadamard
function in the wave-zone regime and use it to calculate  the vacuum
fluctuations and the vacuum expectation for the energy density flux in
the wave zone.

The paper is organized as follows. Section 2 discusses the set up of
the problem. We formulate the equation for a massless scalar field in a
presence of a uniformly accelerated spherical mirror. We also discuss
here Wick's rotation. In Section 3 we obtain a series representation
for the Euclidean Green's function. An analytical continuation to the
Minkowski spacetime and wave-zone regime are discussed in Section 4. In
Sections 5 and 6 we calculate $\la \varphi^2(x)\ra^{\ind{ren}}$ and the
energy density flux   in the wave zone.

\section{Model and Geometry}\label{s2}
\setcounter{equation}0

Our purpose is to study quantum radiation from   a uniformly
accelerated small spherical body with a mirror-like boundary. In order
to describe such a motion it is convenient to introduce in the
Minkowski spacetime with a metric
\be \n{2.1}
ds^2=-dT^2 +dX^2 +dY^2 +dZ^2  \, ,
\ee
new (Rindler) coordinates
\be \n{2.2}
T=\rho \, \sinh \tau \, , \hspace{0.5cm} Z=\rho\, \cosh \tau \, ,
\hspace{0.5cm}X=X\, , \hspace{0.5cm} Y=Y\, .
\ee
These coordinates cover the right wedge $R_+$ of the spacetime where
$Z>|T|$. A line $\rho=\rho_0$, $X=0$, $Y=0$ represent a world line of a
uniformly accelerated observer, moving with a constant acceleration
$w=\rho_0^{-1}$ (as measured in his reference frame) in the
$Z$-direction, while the 3-dimensional plane $\tau$= const is a set of
events which are simultaneous from the point of view of the observer.
In these coordinates metric (\ref{2.1}) is
\be \n{2.3}
ds^2=- \rho^2 d\tau^2 +d\rho^2 +dX^2 +dY^2\, ,
\ee
and the boundary $\Sigma_+$ of a uniformly accelerated spherical mirror is
described by the equation
\be \n{2.4}
X^2 +Y^2 + (\rho -\rho_0)^2 =b^2\, ,
\ee
where $b$ is the radius of the mirror which is smaller than the distance
to the horizon $\rho_0$. This equation in the original
Cartesian coordinates is
\be \n{2.5}
X^2 +Y^2 + (\sqrt{Z^2 -T^2}-w^{-1}))^2 =b^2\, .
\ee
In fact, since (\ref{2.5}) is invariant under the reflection $Z\to -Z$,
it describes two surfaces, $\Sigma_+$ and $\Sigma_-$ which 
correspond to the solution of (\ref{2.5}) with $Z>0$ and $Z<0$,
respectively.

It is evident that for this type of motion, the mirror $\Sigma_+$ can
affect the state of the field only in the region where $T-Z>0$,
while the region lying under the null plane $N_- :
T-Z=0$ is causally disconnected from the region of influence of
$\Sigma_+$.

Our aim is to study the influence of an accelerated motion of
a spherical mirror  on a scalar massless quantum field $\varphi$ . 
In our consideration the
background geometry is flat. Nevertheless,   it is convenient to
consider at first a general action for a scalar massless field in
curved  geometry
\be
W[\varphi ] = -\,\frac{1}{2}\int dx^4\,g^{1/2}\,\left(g^{\mu\nu}\,\varphi_{\!,\mu}\,
\varphi_{\!,\nu} + \xi R\,\varphi^2\right).  \n{2.6}
\ee
The field  equation is
\be
\Box \varphi -  \xi R\,\varphi = 0\,,  \n{2.7}
\ee
where $\Box = g^{-1/2}\,\partial_{\mu}\!\left(g^{1/2}
\,g^{\mu\nu}\partial_{\nu}\right)$. The 
stress-energy tensor, $T_{\mu\nu}$, for the field has the form
\[
T_{\mu\nu} = (1-2\xi )\,\varphi_{\!,\mu}\,\varphi_{\!,\nu} + \left(2\xi - 
\frac{1}{2}\right)
g_{\mu\nu}\,g^{\alpha\beta}\,\varphi_{\!,\alpha}\,\varphi_{\!,\beta }
\]
\be
\hspace{2.9cm} - \,2\xi\left(\varphi \varphi_{;\mu\nu} - 
g_{\mu\nu}\,\varphi_{;\alpha}\,
\varphi^{;\alpha}\right) + \xi \left(R_{\mu\nu} - \frac{1}{2}g_{\mu\nu}\,R\right)
\varphi^2.  \n{2.8}
\ee
When the parameter $\xi$ of non-minimal coupling vanishes, $T_{\mu\nu}$
reduces to  the canonical stress-energy tensor. The case
\be
\xi = 1/6  \n{2.9}
\ee
corresponds to the conformally invariant theory with $T_{\mu}^{\mu} = 0 $.

For our problem, the curvature $R$ vanishes, the field $\varphi$
propagates in the exterior of the surface $\Sigma$, and obeys the
Dirichlet boundary condition on $\Sigma$
\be \n{2.10}
\varphi |_{\Sigma}=0\, .
\ee
We denote by $G^{(1)}(x,x')$ the corresponding Hadamard function
\be \n{2.11}
G^{(1)}(x,x') = \la 0|\hat{\varphi}(x)\,\hat{\varphi}(x') + 
\hat{\varphi}(x')\,\hat{\varphi}(x)|0\ra \,.
\ee
This is symmetric function of its arguments $x$ and $x'$ obeying the
boundary conditions
\be \n{2.12}
G^{(1)}(x,x')|_{x\in\Sigma}=G^{(1)}(x,x')|_{x'\in\Sigma}=0\, .
\ee
Hadamard function depends on the choice of the state of the quantum
field. Since our problem possesses the invariance with respect to a
reflection $T\to -T$, we use the state which also obeys the same
symmetry. This is evidently a preferable state for which the calculations
are greatly simplified. For this state, the Hadamard function
$G^{(1)}(x,x')$ can be obtained by the standard Wick's rotation
prescription. 

Namely, let us make a rotation of time $T$ in the complex plane
\be \n{2.13}
T\to i {\cal T}\, .
\ee
Under this rotation the metric becomes Euclidean
\be \n{2.14}
ds_E^2= d{\cal T}^2 +dX^2 +dY^2 +dZ^2  \, ,
\ee
and the equation of the mirror surface takes the form
\be \n{2.15}
X^2 +Y^2 + (\sqrt{Z^2 +{\cal T}^2}-w^{-1})^2 =b^2\, .
\ee
This surface $\Sigma_E$ is a 4-dimensional torus $S^1\times S^2$, which
is obtained by the rotation of a sphere $S^2$ of the radius $b$ around
a circle $S^1$ of the radius $w^{-1}$ ($b<w^{-1}$).  Denote by
$G_E(x,x')$ the Euclidean Green function, that is a symmetric (with
respect $x$ and $x'$) solution of the equation
\be \n{2.16}
\Box_E G_E(x,x')= -\delta(x'x')\, ,
\ee
defined in the exterior of the torus $\Sigma_E$ and obeying the boundary
conditions
\be \n{2.17}
G_E(x,x')|_{x\in\Sigma}=G_E(x,x')|_{x'\in\Sigma}=0\, .
\ee
For space-like separation of the arguments, the Hadamard function
$G^{(1)}$ can be obtained from $G_E$ by the Wick's rotation
\be \n{2.18}
G^{(1)}(x,x')=2 G_E(x,x')|_{{\cal T}\to -iT,{\cal T}'\to -iT'}\, . 
\ee

It should be emphasized that using the Euclidean approach greatly
simplifies calculations. On the other hand this method has restrictions. In
particular, the Green's functions obtained by the analytical
continuation of $G_E(x,x')$ correspond to the averages of the
$\hat{\varphi}(x)\hat{\varphi}(x')$ for a special quantum state singled
out by its $T\to -T$ invariance. Moreover, because of the symmetry $Z\to
-Z$, the corresponding Green's function obeys the Dirichlet boundary
conditions on the both surfaces, $\Sigma_+$ and $\Sigma _-$, and hence
always gives us a result when two accelerated mirror are present.

In the next sections we shall obtain an expression for the Euclidean
Green function and discuss the problem of its analytical continuation to
the physical Minkowski spacetime later.

\section{Euclidean Green Function}\label{s3}
\setcounter{equation}0

The Euclidean Green function $G_E (x,x')$ coincides with an electric
potential at a point $x$ created by a point charge located at a point
$x'$ of the 4-dimensional space in the presence of a conducting surface
$\Sigma_E$. This problem can be solved by using the toroidal
coordinates, for which the 4-dimensional Laplace operator $\Box_E$
allows separation of variables. The toroidal coordinates $(\eta, \psi,
\gamma, \phi)$ are related to the Cartesian coordinates as follows
\be \n{3.1}
X={a\sin\gamma\over B}\cos \phi\, ,\hspace{0.5cm}
Y={a\sin\gamma\over B}\sin \phi\, ,
\ee
\be \n{3.2}
Z={a\sinh\eta\over B}\cos \psi\, ,\hspace{0.5cm}
{\cal T}={a\sinh\eta\over B}\sin \psi\, ,
\ee
where $B=\cosh\eta -\cos \gamma$. The metric (\ref{2.14}) in these
coordinates takes the form
\be \n{3.3}
ds_E^2=\Omega^2 \, \, d\tilde{s}^2\, ,\hspace{0.5cm}
\Omega={a\over B}\, ,
\ee
\be \n{3.4}
d\tilde{s}^2=dH^2 + dS^2\, ,\hspace{0.5cm}
\ee
where
\be \n{3.5}
dH^2= d\eta^2 +\sinh^2\eta\, d\psi^2\, 
\ee
is a metric on a hyperboloid $H^2$ and
\be \n{3.6}
dS^2= d\gamma^2 +\sin^2\gamma \, d\phi^2\, 
\ee
is a metric on the unit sphere $S^2$. 

In these coordinates surfaces $\eta=$const are torii. By substituting
(\ref{3.1})--(\ref{3.2}) into (\ref{2.15}) one can easily find that
\be\n{3.7}
a=\sqrt{{1\over w^2}-b^2}\, ,\hspace{0.5cm} \tanh \eta_0=\sqrt{1-w^2
b^2}\, ,
\ee
where $\eta_0$ is the value of $\eta$ corresponding to $\Sigma_E$.
Points with $\eta <\eta_0$ lie in the exterior of $\Sigma_E$. 

Later we shall use the following expressions in the toroidal
coordinates for the square of the distance ${\cal R}^2$ from the origin
to the point $x=(X,Y,Z,{\cal T})$ and for the square of the distance
${\cal R}^2(x,x')$ between points $x$ and $x'$
\be\n{3.8}
{\cal R}^2 =a^2 {\cosh \eta +\cos\gamma\over \cosh \eta -\cos\gamma}\, ,
\ee
\be \n{3.9}
{\cal R}^2(x,x')={2\, a^2\, (\cosh \Lambda -\cos\lambda)\over (\cosh
\eta -\cos\gamma)(\cosh \eta' -\cos\gamma')}\, ,
\ee
where $\lambda$ and $\Lambda$ are geodesic distances on a unit sphere
$S^2$ and a unit hyperboloid $H^2$, respectively. They are defined as
\be \n{3.10}
\cos \lambda = \cos \gamma \, \cos \gamma' + \cos (\phi-\phi')\, \sin \gamma
\,\sin \gamma' \, ,
\ee
and
\be \n{3.11}
\cosh \Lambda = \cosh \eta \, \cosh \eta' - \cos (\psi-\psi')\, \sinh
\eta
\,\sinh \eta' \, .
\ee

Suppose we have two conformally related 4 dimensional spaces with
metrics $d\tilde{s}^2$ and $ds^2=\Omega^2 \, d\tilde{s}^2$. Then the
box-operators in these spaces are related as follows
\be \n{3.12}
\Box -{1\over 6}R = \Omega^{-3}\, (\tBox-{1\over 6}\tilde{R}) \,
\, \Omega
\, .
\ee
In our case curvatures $R_E$ and $\tilde{R}$ for metrics $ds_E^2$ and
$d\tilde{s}^2$ given by (\ref{3.3}) -- (\ref{3.4}) vanish and using
(\ref{3.12}) one gets
\be \n{3.13}
\tBox \, \tilde{G} (x,x')=-\tilde{\delta}(x,x')\, ,
\ee
where
\be \n{3.14}
G_E(x,x')=\Omega^{-1}(x)\, \Omega^{-1}(x')\, \tilde{G} (x,x')\, ,
\ee
and 
\be \n{3.15}
\tilde{\delta}(x,x') = {\delta(x-x')\over \sqrt{\tilde{g}}}\, .
\ee
The operator $\tBox$ is of the form
\be \n{3.16}
\tBox=\Delta_H +\Delta_S\, ,\hspace{0.5cm}
\ee
where
\be \n{3.17}
\Delta_H = \partial_{\eta}^2 +\coth \eta \, \, \partial_{\eta} +{1\over
\sinh^2{\eta}}\, \, \partial_{\psi}^2 \, ,
\ee
and
\be \n{3.18}
\Delta_S = \partial_{\gamma}^2 +\cot \gamma \, \partial_{\gamma} +{1\over
\sin^2{\gamma}}\,\, \partial_{\phi}^2 \, ,
\ee
are the Laplace operators on the unit hyperboloid and unit sphere,
respectively.

Expanding the Green function $\tilde{G}$ over spherical harmonics
$Y_{\ell m}$ which form a complete set on the unit sphere we can write
\be \n{3.19}
\tilde{G} (x,x')=\sum_{\ell=0}^{\infty}\, \sum_{m=-\ell}^{\ell}\,
\tilde{G}_{\ell}(p,p')\, Y_{\ell m}(q)\, \bar{Y}_{\ell m}(q')
={1\over 4\pi} \sum_{\ell=0}^{\infty}\, (2\ell+1)\,
\tilde{G}_{\ell}(p,p')\, P_{\ell}(\cos \lambda)\, ,
\ee
where $x=(p,q)$, $x'=(p',q')$, and $p$, $p'$ are points on $H$ and $q$,
$q'$ are points on $S$. Here $P_{\ell}(z)$ is the Legendre polynomial
and $\lambda$ is a geodesic distance (angle) between $q=(\gamma,\phi)$ 
and $q'=(\gamma',\phi')$ on $S$ defined by relation (\ref{3.10}).

The functions $\tilde{G}_{\ell}(p,p')$ are 2-dimensional Green functions
of the operator $\Delta_H -\ell (\ell +1)$ which are regular inside the
disc $0\le\eta <\eta_0$ and obey the Dirichlet boundary conditions at the
boundary of the disc.

Using the Fourier decomposition with respect to the angle variable
$\psi$ one get the following representation for $\tilde{G}_{\ell}(p,p')$
\be \n{3.20}
\tilde{G}_{\ell}(p,p') ={1\over 2\pi}\, 
\sum_{m=-\infty}^{\infty} \, e^{-im(\psi-\psi')}\, {\cal G}_{\ell m}(\eta,\eta')\, ,
\ee
where ${\cal G}_{\ell m}(\eta,\eta')$ obeys the equation
\be \n{3.21}
\left[{d^2\over d\eta^2}+\coth \eta {d\over d\eta} -{m^2\over
\sinh^2\eta} -\ell(\ell +1)\right] {\cal G}_{\ell m}(\eta,\eta')=
-{\delta(\eta-\eta') \over \sinh\eta}\, ,
\ee
and the boundary condition 
\be \n{3.22}
{\cal G}_{\ell m}(\eta_0,\eta')={\cal G}_{\ell m}(\eta,\eta_0)=0\, .
\ee
The required Green functions ${\cal G}_{\ell m}$ must be also regular at
$\eta=0$.

Linear independent solutions of the homogeneous version of the equation
(\ref{3.21}) are the associated  Legendre functions $P_{\ell}^{m}$ and
$Q_{\ell}^{m}$. For complex values of their argument $z$ and parameters
$\nu, \mu$ these functions are defined as (see \cite{BeEr}, eq. 3.2.3
and 3.2.5)
\be \n{3.23}
P_{\nu}^{\mu}(z) =
{1\over \Gamma(1-\mu)}\, \left({z+1\over z -1}\right)^{\mu/2}\, 
F(-\nu, \nu+1; 1-\mu;(1-z)/2)\, ,
\ee
\[
Q_{\nu}^{\mu}(z)= e^{i\mu\pi} \, 2^{-\nu
-1}\sqrt{\pi}{\Gamma(\nu+\mu+1)\over \Gamma(\nu +3/2)}\, z^{-\nu-\mu-1}\,
(z^2-1)^{\mu/2}\,
\]
\be \n{3.24}
\times F(1+{\nu\over 2}+{\mu\over 2}, {1\over 2}+{\nu\over
2}+{\mu\over 2}; \nu +{3\over 2}; z^{-2})\, .
\ee
The function $F$ which enters these relations is the hypergeometric function

Since $\ell$ and $m$ are independent parameters in the
equation, we shall need the Legendre functions both for $|m|\le \ell$
and for $|m|> \ell$. For the latter case and for the standard definition
of $P_{\ell}^{m}(z)$ these functions vanish, while
$\Gamma(\ell-m+1)P_{\ell}^{m}$ remains finite in the limit of integer
$\ell$ and $m$. Thus instead of $P_{\nu}^{\mu}(z)$ it is more
convenient to  use the following functions
\be \n{3.25}
{\cal P}_{\nu}^{\mu}(z)=\Gamma(\nu-\mu+1)P_{\nu}^{\mu}(z) \, .
\ee
It is understood, that for integer $\nu$ and $\mu$ these functions are
defined by continuity. Functions ${\cal P}_{\ell}^m$ are regular at $z=1$. 
We also define
\be \n{3.26}
{\cal Q}_{\nu}^{\mu}(z)={e^{-\mu\pi}\over \Gamma(\nu+\mu+1)}\,
Q_{\nu}^{\mu}(z)\, .
\ee
Functions ${\cal P}_{\nu}^{\mu}(z)$ and ${\cal Q}_{\nu}^{\mu}(z)$
are analytic functions of their  complex arguments
$\nu$, $\mu$, and $z$ defined in the complex plane $z$ with a cut along
the real axis lying to the left of $z=1$. For integer value  $\mu=m$
these functions obey the following symmetry relations
\be \n{3.27}
{\cal P}_{\nu}^{-m}(z)={\cal P}_{\nu}^{m}(z)\, ,\hspace{0.5cm}
{\cal Q}_{\nu}^{-m}(z)={\cal Q}_{\nu}^{m}(z)\, .
\ee
Using relation (3.2.13) of \cite{BeEr}  we obtain  the following
expression for the Wronskian of the functions ${\cal P}_{\nu}^{\mu} (z)$ and 
${\cal Q}_{\nu}^{\mu} (z)$
\be \n{3.28}
W[{\cal P}_{\nu}^{\mu} (z),{\cal Q}_{\nu}^{\mu} (z)]\equiv {\cal P}_{\nu}^{\mu}(z)
{d\over dz}{\cal Q}_{\nu}^{\mu} (z)
-{\cal Q}_{\nu}^{\mu} (z){d\over dz}{\cal P}_{\nu}^{\mu}(z)={1\over
1-z^2}\, .
\ee

The following functions are vanishing at $z=z_0$
\be\n{3.29}
{\cal O}_{\ell}^m (z|z_0)={\cal Q}_{\ell}^m (z)-{ {\cal Q}_{\ell}^m (z_0)
\over {\cal P}_{\ell}^m (z_0)}\, {\cal P}_{\ell}^m (z)\, .
\ee
The Green function ${\cal G}_{\ell m}(\eta,\eta')$ is
\be \n{3.30}
{\cal G}_{\ell m}(\eta,\eta')=\, {\cal P}_{\ell}^m
(\cosh \eta_{<})\, {\cal O}_{\ell}^m (\cosh \eta_{>}|z_0)\, ,
\ee
where $z_0=\cosh \eta_0$. 
Combining the obtained results we get the following representation for the
Euclidean Green function $\tilde{G}$
\be \n{3.31}
\tilde{G} (x,x')
={1\over 8\pi^2} \sum_{\ell=0}^{\infty}\, (2\ell+1)\,
 P_{\ell}(\cos \lambda) \, \sum_{m=0}^{\infty} \, \, 
\beta_m\, \cos[m(\psi-\psi')]\, {\cal P}_{\ell}^m
(\cosh \eta_{<})\, {\cal O}_{\ell}^m (\cosh \eta_{>}|z_0)\, ,
\ee
where $\beta_0=1$ and $\beta_{m\ge 1}=2$. In order to get this relation
we used the symmetry properties (\ref{3.27}).

Function ${\cal Q}_{\ell}^m (z)$ which enter the definition (\ref{3.29})
of ${\cal O}_{\ell}^m (z|z_0)$ evidently does not depend on on the
position of the boundary $z_0$. Let us calculate now the corresponding
boundary independent part of $\tilde{G} (x,x')$. Namely, we denote by 
$\tilde{G}^0(x,x')$ the expression (\ref{3.31}) where instead of ${\cal
O}_{\ell}^m$ we substitute its boundary independent part ${\cal
Q}_{\ell}^m$.  Using relation 3.11.4 from \cite{BeEr} we have
\be\n{3.32}
\sum_{m=0}^{\infty} \, \, 
\beta_m\, \cos[m(\psi-\psi')]\, {\cal P}_{\ell}^m
(z_{<})\, {\cal Q}_{\ell}^m (z_{>})= Q_{\ell}( z_< z_>
-\cos(\psi-\psi')\sqrt{ (z_>^2-1)(z_<^2-1)})  \, .
\ee
Using this relation together with Heine formula (see equation 3.11.10 of
Ref.\cite{BeEr})
\be \n{3.33}
\sum_{\ell=0}^{\infty}\, (2\ell+1)\, P_{\ell}(t') Q_{\ell}(t)={1\over
t-t'}\, 
\ee
we obtain
\be\n{3.34}
\tilde{G}^0(x,x')= {(\cosh
\eta -\cos\gamma)(\cosh \eta' -\cos\gamma')\over 8\pi^2\, a^2\, 
(\cosh \Lambda -\cos\lambda)}\, ,
\ee
where $\lambda$ and $\Lambda$ are defined by (\ref{3.10})and (\ref{3.11}).
This result implies that $G_E^0$ related to $\tilde{G}^0$ by
(\ref{3.14}) coincides with the {\em vacuum} Green function
\be\n{3.35}
G_E^0(x,x')={1\over 4\pi^2 {\cal R}^2(x,x')}\, .
\ee

Thus the renormalized Euclidean Green defined as
\be \n{3.36}
G_E^{\ind{ren}}(x,x')=G_E(x,x')- G_E^0(x,x')
\ee
has the following series representation
\[
 G_E^{\ind{ren}}(x,x')
=-{B B'\over 8\pi^2 a^2} \sum_{\ell=0}^{\infty}\, (2\ell+1)\,
 P_{\ell}(\cos \lambda) \, \sum_{m=0}^{\infty} \, \, 
\beta_m\, \cos[m(\psi-\psi')]\, 
\]
\be \n{3.37}
{ }\quad \quad \quad \times\,{\cal P}_{\ell}^m(\cosh \eta)\, 
{\cal P}_{\ell}^m (\cosh \eta')\,  {  {\cal Q}_{\ell}^m (\cosh \eta_0)
\over {\cal P}_{\ell}^m (\cosh \eta_0)} \, , 
\ee
where $B=\cosh \eta -\cos \gamma$ and $B'=\cosh \eta' -\cos \gamma'$.

\section{Wave Zone Regime}\label{s4}
\setcounter{equation}0

We discuss now the problem of analytical continuation of the results
obtained in the Euclidean space to the ``physical'' spacetime with the
Minkowski metric. In order to do this it is sufficient to express
coordinates $(\eta, \psi, \gamma, \phi)$ in terms of Euclidean
coordinates $({\cal T}, X, Y, Z)$ and after make the Wick's rotation
${\cal T} \to -iT$. To simplify calculations we  assume that the
arguments of $ G_E^{\ind{ren}}$ are split in time direction. Namely
this object is of interest for our calculations. Thus we have 
\be
 G_E^{\ind{ren}}({\cal X},{\cal X}')
=-{B B'\over 8\pi^2 a^2} \sum_{\ell=0}^{\infty}\, (2\ell+1)\,
\, \sum_{m=0}^{\infty} \, \, 
\beta_m\, 
\,{\cal P}_{\ell}^m(\cosh \eta)\, 
{\cal P}_{\ell}^m (\cosh \eta')\,  {  {\cal Q}_{\ell}^m ((wb)^{-1})
\over {\cal P}_{\ell}^m ((wb)^{-1})} \, , 
\ee
where ${\cal X}=({\cal T},X,Y,Z)$ and ${\cal X}'=({\cal T}',X,Y,Z)$.
We used also that $\cosh \eta_0 =(wb)^{-1}$. Simple calculations also
give
\be
{B^2\over a^2}={4a^2 \over ({\cal R}^2-a^2)^2+4a^2(X^2+Y^2)}\, ,
\ee
\be
\cosh^2 \eta ={ ({\cal R}^2+a^2)^2\over ({\cal
R}^2-a^2)^2+4a^2(X^2+Y^2)}\, ,
\ee
where ${\cal R}^2={\cal T}^2+ R^2$ and $R^2=X^2+Y^2 +Z^2$.

If we are interested in the wave-zone limit of $G_E^{\ind{ren}}$ it can
be further simplified. This limit can be found if one makes the Wick's
rotation ${\cal T} \to -iT$ first, then puts 
\be
T=U+R\, ,\hspace{0.5cm} X=R \sin \Theta \cos \Phi \, ,\hspace{0.5cm}
Y=R \sin \Theta \sin \Phi \, ,\hspace{0.5cm}
Z=R \cos \Theta\, 
\ee
and after this takes the limit $R\to \infty$ with $(U,\Theta,\Phi)$
fixed. Keeping the leading terms one gets
\be
{B^2\over a^2}={a^2 z^2\over R^2\, U^2}\, ,
\ee
\be
\cosh \eta =z\, ,\hspace{0.5cm}
z=
{1\over \sqrt{1+a^2\sin^2\Theta/U^2} }\, .
\ee
The right-hand side of the expression for $\cosh^2 \eta$ is less 1.
This means that as a result of Wick's rotation $\eta$ takes pure
imagine value. In the wave zone we have 
\be\n{4.7}
 G_E^{\ind{ren}}(U,U';R, \Theta)\sim {{\cal G}(U,U';\Theta)\over R\, R'}\, ,
\ee
\be\n{4.8}
{\cal G}(U,U';\Theta)=-{a^2 z z'\over 8\pi^2\,U\,U'}
\sum_{\ell=0}^{\infty}\, (2\ell+1)\,
\, \sum_{m=0}^{\infty} \, \, 
\beta_m\, 
\,{\cal P}_{\ell}^m(z)\, 
{\cal P}_{\ell}^m (z')\,  {  {\cal Q}_{\ell}^m (z_0)
\over {\cal P}_{\ell}^m (z_0} \, , 
\ee
where $z_0=(wb)^{-1}$.

\section{$\la \varphi^2\ra^{\ind{ren}}$ in the Wave Zone}
\setcounter{equation}0

In the coincidence limit the function $ G_E^{\ind{ren}}$ is finite
and  gives $\la \varphi^2\ra^{\ind{ren}}$. Thus we have
 the following representation for 
$\la \varphi^2\ra^{\ind{ren}}$  in the wave zone
\be\n{5.5}
\la \varphi^2(x)\ra^{\ind{ren}}= - {a^2 \over 8\pi^2 R^2 U^2}\,
{\cal F}(z,z_0)\, ,
\ee
where
\be \n{5.6}
{\cal F}(z,z_0)= z^2 \, \sum_{\ell=0}^{\infty}\, (2\ell+1)\,
\, \sum_{m=0}^{\infty} \, \, \beta_m\, \,[{\cal P}_{\ell}^m(z)]^2\, 
\,  {  {\cal Q}_{\ell}^m (z_0) \over {\cal P}_{\ell}^m (z_0)} \, ,
\ee
and
\be\n{5.7}
z={1\over \sqrt{1+(a\sin \Theta/U)^2}}\, ,\hspace{0.5cm}
a=\sqrt{{1\over w^2}-b^2}\, .
\ee
The function ${\cal F}(z,z_0)$ depends on constants $w$ and $b$ which
specify the problem. Besides this it depends on the coordinates $U$ and
$\Theta$ on ${\cal J}^+$ which enter only through a combination
$U/\sin\Theta$. The independence of the result of $\Phi$ is a
consequence of the invariance of the mirror-like boundary $\Sigma$,
(\ref{2.5}), under rotation in the $X$--$Y$ -plane. Moreover, the
equation (\ref{2.5}) for $\Sigma$ is also invariant under a boost
transformation in the $T$--$Z$-plane. It can be shown  (see e.g. Ref.
\cite{FrSi:00}) that as the result of this symmetry $\la
\varphi^2\ra^{\ind{ren}}$ near ${\cal J}^+$ must have the form $\sim
R^{-2} U^{-2} f(\sin\Theta/U)$. The fact that our result (\ref{5.5})
does have the form dictated by the symmetry of the problem gives an
independent test of the correctness of the calculations. It is also
easy to see that all the dependence on $U$ enters through the
dimensionless time parameter $u=U/a$.

If the size of the ball is small
we have $z_0\to\infty$. In this regime one can use the following
asymptotics
\be\n{5.8}
{\cal P}^{\mu}_{\nu}(z_0)\sim {(2z_0)^{\nu}\over \sqrt{\pi}}\,
\Gamma(\nu+{1\over 2})\, ,\hspace{0.5cm} Re(\nu)> -1/2\, ,
\ee
\be\n{5.9}
{\cal Q}^{\mu}_{\nu}(z_0)\sim {(2z_0)^{-\nu-1}\, \sqrt{\pi}\over \,
\Gamma(\nu+{3\over 2})}\, ,
\ee
and hence
\be\n{5.10}
{{\cal Q}^{\mu}_{\nu}(z_0)\over {\cal P}^{\mu}_{\nu}(z_0)}\sim 
{2\pi \over (2z_0)^{2\nu+1}(2\nu+1)\,
[\Gamma(\nu+{1\over 2})]^2}\, ,\hspace{0.5cm}Re(\nu)> -1/2\, .
\ee
Since the asymptotic of this ratio does depend on $m$ we can write the
small $b$ expansion of ${\cal F}(z,z_0)$ as follows
\be \n{5.11}
{\cal F}(z,z_0) =2\pi z^2 \sum_{\ell =0}^{\infty} {1 \over (2z_0)^{2\ell+1}\,
[\Gamma(\ell+{1\over 2})]^2}\, F_{\ell}(z)\, ,
\ee
where 
\be \n{5.12}
F_{\ell}(z) =\sum_{m=0}^{\infty}\, \beta_m\, [{\cal P}^m_{\ell}(z)]^2\, .
\ee

The leading (proportional to $b$) contribution for small
$b$ is given by $\ell=0$ term in the series (\ref{5.12}). 
Notice that for integer $m$
\be \n{5.13}
{\cal P}^m_0 (z)= 2^{-m} (z^2-1)^{m/2}\, F(m+1,m;m+1;{1-z \over 2})\, ,
\ee
(see formula 3.6.1.1 of \cite{BeEr}). Using the following property of
the hypergeometric function
\be \n{5.14}
F(b,a;b;\xi)=(1-\xi)^{-a}\, ,
\ee
we get
\be\n{5.15}
{\cal P}^m_0 (z)= \left( {z-1\over z+1}\right)^{m/2}\, .
\ee

To calculate ${\cal P}^m_{\ell} (z)$  for $\ell >1$
 one can use the following relation
\be\n{5.16}
{\cal P}^m_{\ell+1}(z)=(z^2-1){d{\cal P}^m_{\ell}(z)\over dz}+(\ell +1)\, z\, 
{\cal P}^m_{\ell}(z)\, .
\ee
In particular, one has
\[
{\cal P}^m_{1}(z)=(z+m)\, \left( {z-1\over z+1}\right)^{m/2}\,
,
\]
\be\n{5.17}
{\cal P}^m_{2}(z)=(3z^2+3zm+m^2-1)\, \left( {z-1\over z+1}\right)^{m/2}\,
.
\ee

Summation of series (5.12) can be performed in the closed form by using
Maple program. It can be shown that $F_{\ell}(z)$ is a polynomial of $z$
of the order $2\ell +1$. The first few harmonics of $F_{\ell}(z)$ are
\[
F_0(z) =z\, ,
\]
\be\n{5.18}
F_1(z) ={1\over 2}(5z^3-2z^2-3z+2)\, ,
\ee
\[
F_2(z) ={z\over 2}(63z^4-18z^3-70z^2+12z+15)\, .
\]

Using these results one can show that the leading contribution to $\la
\varphi^2(x)\ra^{\ind{ren}}$ for the small radius of 
the mirror $b\ll w^{-1}$ is
\be \n{5.19}
\la \varphi^2(x)\ra^{\ind{ren}}\sim -{bw\over 8\pi^2 R^2 u^2\, 
[1+(\sin\Theta/u)^2]^{3/2}}\,  ,
\ee
where $u=U/a\approx wU$ is the dimensionless retarded time. By using
relations (\ref{5.5}), (\ref{5.11}), and (\ref{5.18}) one can also
easily obtain higher corrections to $\la \varphi^2(x)\ra^{\ind{ren}}$
as powers of  $bw$.

\section{Energy Density Flux in the Wave Zone}
\setcounter{equation}0

The calculation of the energy density flux in the solid angle $d\Omega$ 
in the wave zone 
\be \n{6.1}
{dE \over dU d\Omega} = \lim_{R\to \infty}  R^2 \, \la T_{UU} \ra^{\ind{ren}}\, 
\ee
is similar to the calculation of $\la \varphi^2(x)\ra^{\ind{ren}}$ but
more involved.

Using (\ref{4.8}) it is easy to show that when one calculates 
$\la T_{UU} \ra^{\ind{ren}}$ it is sufficient to keep only derivatives
with respect to $U$ and $U'$. All other derivatives effectively
introduce extra power of $R^{-1}$ and hence do not contribute to the
flux of the energy density at infinity. Hence in the wave zone we have
\be\n{6.3}
{dE \over dU} =  [(1-2\xi)\partial_U\partial_{U'} -\xi
(\partial_U^2+\partial_{U'}^2)] \, \, {\cal G}(U,U';\Theta)\, ,
\ee
where
\be\n{6.4}
{\cal G}(U,U';\Theta) =-{a^2\over 8\pi^2}{{\cal F}(z,z'|z_0)\over U U'}\, , 
\ee
\be\n{6.5}
{\cal F}(z,z'|z_0)= z z' \, \sum_{\ell=0}^{\infty} (2\ell +1) {\cal
F}_{\ell}(z,z'|z_0)\, ,
\ee
\be \n{6.6}
{\cal F}_{\ell}(z,z'|z_0)=\sum_{m=0}^{\infty}\beta_m 
{\cal P}^m_{\ell}(z)\, {\cal P}^m_{\ell}(z')\, {  {\cal Q}_{\ell}^m (z_0)
\over {\cal P}_{\ell}^m (z_0)}\, .
\ee
In the lowest order in $bw$ one has
\be
{\cal F}(z,z'|z_0)\sim {z z'\over z_0}\, {
\sqrt{(1+z)(1+z')}+\sqrt{(1-z)(1-z')}\over
\sqrt{(1+z)(1+z')}-\sqrt{(1-z)(1-z')}}\, .
\ee
To obtain this result we took into account that only $\ell=0$
contributes in this order, and used expression (\ref{5.15}).

It is easy to check that
\be\n{6.7}
{\partial z\over \partial U}={z(1-z^2)\over U}\, .
\ee
Thus we have
\be\n{6.8}
\partial_U = \partial_U|_z +{z(1-z^2)\over U}\,\partial_z|_U\, .
\ee
We use notation $\partial_a|_b$ to indicate that the partial derivative
with respect to $a$ is taken by assuming that $b$ is fixed. Direct
calculations by using Maple give
\be\n{6.9}
{dE\over dU}\sim -{a^2 \over 8\pi^2 z_0 U^4}\, f(z,\xi)\, ,
\ee
where
\be\n{6.10}
f(z,\xi)=z\left[(1+z^2)-2\xi(2+z^2+3z^4)\right]\, .
\ee
Since $a\sim w^{-1}$, $z_0\sim (wb)^{-1}$ we have
\be\n{6.11}
{dE\over dU}\sim -{b w^3 \over 8\pi^2 u^4} \, f(z,\xi)\, .
\ee
For small values of $u$ and $\Theta \ne 0, \pi$, $z\sim u/|\sin \Theta|$
and hence 
\be
{dE\over dU}\sim -{bw^3(1-2\xi)\over 8\pi^2|\sin \Theta| u^3}\, .
\ee
 For large $u$ one has
\be
{dE\over dU}\sim -{bw^3(1-6\xi)\over 8\pi^2 u^4}\,. 
\ee

\section{Conclusion}

In this paper we considered the vacuum polarization effects in the
presence of  uniformly accelerated spherical mirrors. We considered a
scalar field model. Under assumption that a size of the mirror, $b$, is
much smaller than the inverse acceleration, $w^{-1}$, we calculated the
field fluctuation and the energy density flux created by the
accelerated body at infinity. This flux is given by (\ref{6.11}), and
for the canonical energy (i.e. for $\xi=0$) it is always negative. Its
divergence $\sim U^{-3}$ near $U=0$ is connected with the idealization
of the problem: it is assumed that the motion remains uniformly
accelerated for an infinite interval of time. The boost-invariance
property connected with this assumption significantly simplifies
calculations. In particular, the quantity $dE/(dU d\Omega)$, which
describes the angular distribution of the energy density flux at ${\cal
J}^+$ at given moment of the retarded time $U$, besides common scale
dependence, $U^{-4}$, depends on  one invariant variable,
$\sin\Theta/aU$. 

It should be emphasized again that due to the symmetry of the problem
the Euclidean approach used in this paper gives a result for two-mirror
system, one of them moving in $R_+$ domain and the other moving in
$R_-$ domain. Thus what we obtain as the result of calculations is
joint radiation of such two mirrors. Moreover, the analytical
continuation of the Euclidean expressions gives the result for the
"vacuum" expectation values for a very special choice of the "vacuum"
state, namely the state which is invariant under boost
transformations.   It would be interesting to study a similar effect of
quantum radiation by a single accelerated mirror and for other quantum
states, e.g. for a state obtained from a Minkowski vacuum by
switching-on procedure. This kind of problems must be investigated
directly in the Minkowski spacetime. Nevertheless, some elements of a
construction used in this paper and connected with the boost symmetry
may be useful.

\bigskip

\vspace{12pt}
{\bf Acknowledgments}:\ \  This work was  partly supported  by  the
Natural Sciences and Engineering Research Council of Canada. One of the
authors (V.F.) thanks the Tokyo Institute of Technology for its
hospitality during the preparation of this paper for publication.

\end{document}